\documentclass[12pt]{iopart}

\usepackage[english]{babel}
\usepackage{graphicx}
\usepackage{cite}
\usepackage{subfigure}

\newcommand{\D}{{\rm d}}

\begin{document}

\title[Resetting in arbitrary dimension]{Diffusion with resetting in arbitrary spatial dimension}
\author{Martin R. Evans$^{(1)}$ and Satya N. Majumdar$^{(2)}$}
\address{$^{(1)}$ SUPA, School of Physics and Astronomy, University of
Edinburgh, Mayfield Road, Edinburgh EH9 3JZ, United Kingdom\\
$^{(2)}$ Univ. Paris-Sud, CNRS, LPTMS, UMR 8626, Orsay F-01405, France}

\begin{abstract}
We consider diffusion in arbitrary spatial dimension $d$ with the
addition of a resetting process wherein the diffusive particle
stochastically resets to a fixed position at a constant rate $r$. We
compute the nonequilibrium stationary state which exhibits
non-Gaussian behaviour.  We then consider the presence of an absorbing
target centred at the origin and compute the survival probability and
mean time to absorption of the diffusive particle by the target. The
mean absorption time is finite and has a minimum value at an optimal
resetting rate $r^∗$. Finally we consider the problem of a finite
density of diffusive particles, each resetting to its own initial
position.  While the typical survival probability of the target at the
origin decays exponentially with time regardless of spatial dimension,
the average survival probability decays asymptotically as $ \exp(-A
(\ln t)^d)$ where $A$ is a constant.  We explain these findings using
an interpretation as a renewal process and arguments invoking extreme
value statistics.
\end{abstract}


\maketitle

\section{Introduction}

Diffusion is a fundamental dynamical process dating back to Einstein
and the properties of the diffusion equation have been much
studied. In this work we consider the problem of diffusion with a
stochastic resetting process whereby the diffusive particle is reset
to a fixed position (which we usually take to coincide with the particle's initial  position) at random times and the diffusive process
begins anew. It turns out \cite{EM1} that resetting fundamentally
affects the properties of diffusion.

Resetting holds the system away from any equilibrium state by
constantly returning the system to the initial condition. It is thus a
simple way of generating a nonequilibrium stationary state. In such
states probability currents are non-zero and detailed balance does not
hold---for resetting systems a non-vanishing steady-state current is directed towards the resetting position.  The nature and properties of nonequilibrium
stationary state are questions of fundamental importance within
statistical physics \cite{KRB}.  Thus, the resetting paradigm provides a convenient
framework within which to study such nonequilibrium properties.

Resetting was first considered in a stochastic multiplicative
model of population growth where stochastic resetting events of the
population size was shown to lead to a stationary distribution in
which the population size has a power-law distribution ~\cite{MZ}.  A
continuous-time random walk model in the presence of a constant drift
and resetting has also been studied and the stationary distribution
established~\cite{MV}.  Related models of population growth include
those involving catastrophic events in which the population is
stochastically reduced and reset to some value lower value
\cite{VAME}.

The resetting paradigm has a natural realisation in the context of
search processes \cite{BenichouRV,Bell}. The optimal stochastic search appears as a classic problem in areas as diverse as computer science \cite{MZ02} (e.g. searching for an element in an array) through biochemistry
\cite{Delbruck}
 (e.g. a protein searching a
binding site) to macrobiology
\cite{Bartumeus}
 (e.g. a predator seeking its prey) .  It
is also a teasing question in everyday life---how best does one search
for lost keys?  One general class of search strategies are termed {\it
  intermittent} and combine periods of slow, local motion, termed
foraging, in which the target may be detected with periods of fast
motion, termed relocation, during which the searcher relocates to
new territory (see \cite{BenichouRV} for a recent review). A diffusive
process mimics the foraging phase and a resetting process mimics the
relocation.  Thus diffusion with stochastic resetting provides a
simple realisation  of an intermittent search process.  
Another related search process introduced in  \cite{Gelenbe} is one in which an individual has  a random lifetime and when the searcher dies, a new searcher is
introduced into the system at the initial starting point.
In the mathematical literature, the mean hitting  time for a class
of random walks in which the walker  may choose to restart the walk has been studied recently from an algorithmic point of view~\cite{JP}.  

In this paper we consider diffusion with stochastic resetting as  a fundamental
nonequilibrium process in which the statistics of first passage
properties may be computed exactly.  A simple model of diffusion with
stochastic resetting, in which a Brownian particle is stochastically
reset to its initial position with a constant rate $r$ was defined and
studied in \cite{EM1} with the focus on one spatial dimension.  Amongst the results obtained,  it was shown in
\cite{EM1} that there exists an optimal resetting rate $r^{*}$ that
minimizes the average hitting time to the target.
Extensions to space-dependent resetting rate, resetting to a random
position with a given distribution and to a spatial distribution
of the target were considered in~\cite{EM2}.
How the average absorption time is increased when the searcher is only partially absorbed by the target, corresponding to an imperfect searcher, has been studied in~\cite{WEM}.
Finally a comparison between the statistics of first passage times for diffusion with stochastic resetting and for   an equilibrium dynamics that generates the same stationary state has been made~\cite{EMM}.

In the present work we extend these studies to diffusion in arbitrary spatial dimension $d$. 
In Section 2 we define the model and solve the forward master equation for the probability distribution of the diffusive particle both in the stationary state
and relaxing to the stationary state. In Section 3 we compute the survival probability  of a fixed absorbing target (or trap) in the presence of a diffusing particle with resetting. We also present a simple interpretation of the result in terms of the extreme value statistics of a renewal process. 
In Section 4 we compute the optimal mean first absorption time in arbitrary
dimension.
In Section 5 we consider
many diffusive particles in the presence of a single trap and study the average and typical behaviours of the survival probability of the trap.
The system furnishes  one of the few models for which all these properties
can be computed exactly in arbitrary dimensions.
Finally we conclude in Section 6 with a summary and outlook.

\section{Diffusive resetting problem in arbitrary spatial dimension}
\subsection{Model definition}
First let us define diffusion with resetting in arbitrary spatial dimension $d$.
We consider a single particle (or searcher)
in $\mathcal{ R}^d$ with
initial position $\vec x_0$ at $t=0$ and resetting to position $\vec X_r$.
We stress here that the initial position  $\vec x_0$ and resetting position
$\vec X_r$ are in general distinct, although at the end of some calculations it is convenient to set them to be equal. Our notation for the resetting position thus differs from that of \cite{EM1}.

\begin{figure}[ht]
  \begin{center}
    \includegraphics[width=0.8\textwidth]{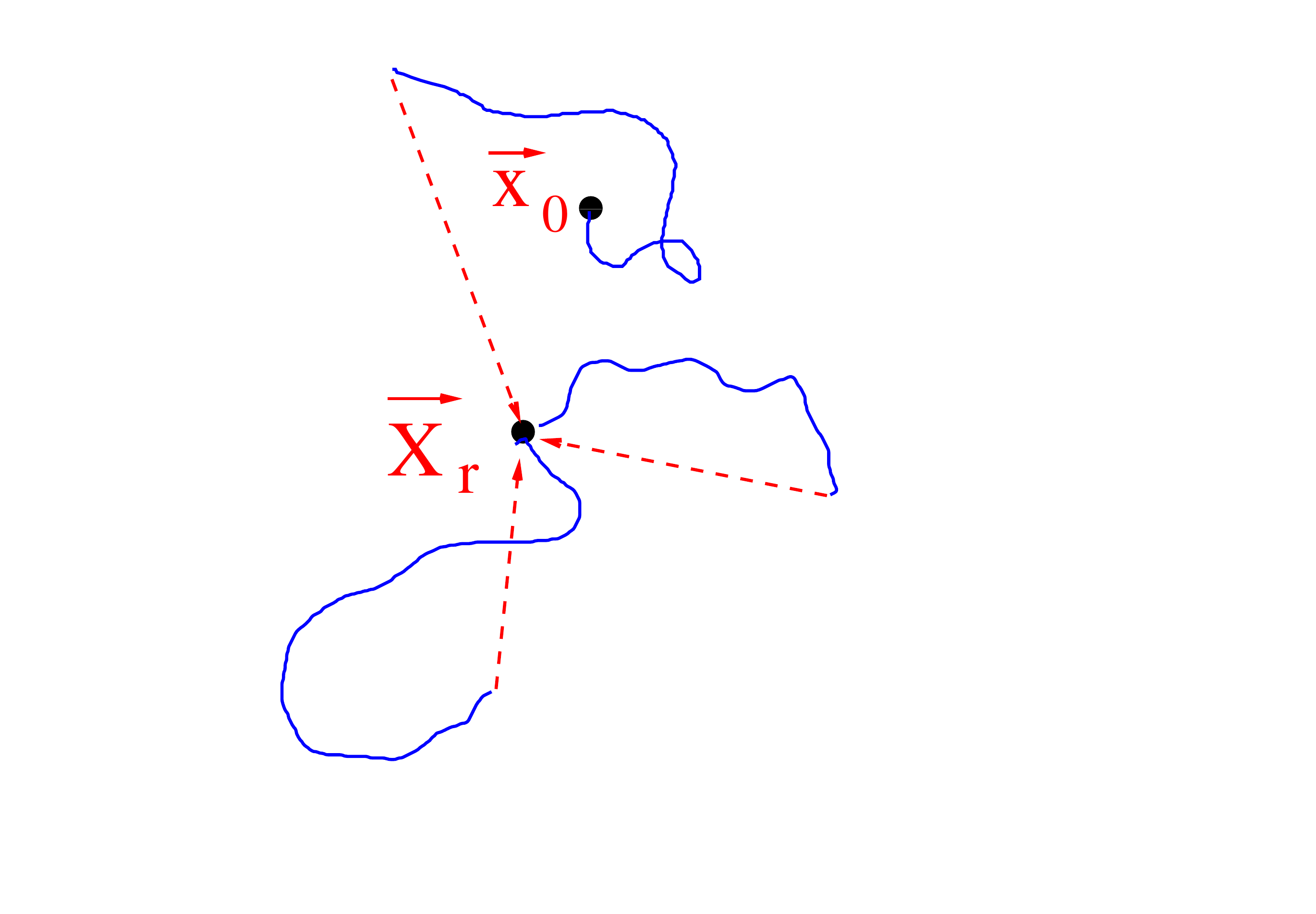}
  \end{center}
  \caption{Illustration in $d=2$ of the diffusion with resetting process: the particle starts at initial position $\vec x_0$ and resets to  position $\vec X_r$ with rate $r$.}
    \label{fig1}
  \end{figure}

The position $\vec x(t)$ of the particle at time
$t$ is updated by the following stochastic
rule~\cite{EM1}: in a small time interval $\D t$ each component $x_i$ of
the position vector $\vec x(t)$ becomes 
\begin{eqnarray}
 x_i(t+dt) & =& ( X_r)_i \quad {\rm with\,\,probability}\,\, r\, dt \nonumber \\
& =&  x_i(t) + \xi_i(t) dt \quad {\rm with\,\,probability}\,\, (1-r\, dt)
\label{rule.1}
\end{eqnarray}
where $\xi_i(t)$ is a Gaussian white noise with mean $\langle \xi_i(t)\rangle =0$
and the two-point correlator $\langle \xi_i(t)\xi_j(t')\rangle= 2\,D\,\delta_{ij} \delta(t-t')$.
The dynamics thus  consists of a stochastic mixture of resetting to
the initial position with rate $r$ (long range move) and ordinary diffusion 
(local  move) with diffusion constant $D$ (see Fig. (\ref{fig1})). 

\subsection{Solution of Master Equation for diffusion with resetting}

The probability density for the particle to be at position $\vec x$ at time $t$, having started from position $\vec x_0$ at time $t=0$ with resetting to position $\vec X_r$, should, in principle,  be written as 
$p(\vec{x},t|\vec{x}_0; \vec X_r)$. However in the following, in order to lighten the notation, we shall suppress the initial and resetting positions and abbreviate to $p(\vec{x},t)$.

The forward master equation for the probability density for diffusion with resetting rate $r$ to point $\vec X_r$ reads
\begin{equation}
\frac{\partial p(\vec{x},t)}{\partial t}
= D\nabla^2p(\vec x,t) - r p(\vec x,t) + r\delta^d(\vec x-\vec X_r)\;,
\label{mehd}
\end{equation}
with initial condition $p(\vec x,0) = \delta^d(\vec x-\vec x_0)$. The first term on the right hand side (r.h.s) of Equation (\ref{mehd}) expresses the diffusive spread of probability;
the second term expresses the loss of probability from $\vec x$ due to resetting to $\vec X_r$; the final term corresponds to the gain of probability at $\vec X_r$ due to resetting from all other positions.

One can write down the solution to (\ref{mehd}) in a simple and intuitive way as follows. We first note that the (initial value) diffusive Green function
in the absence of resetting ($r=0$),   which we denote  $G(\vec{x},t|\vec{x}_0)$, satisfies
\begin{equation}
\frac{\partial G(\vec{x},t|\vec{x_0})}{\partial t} = D\nabla^2 G(\vec
x,t|\vec x_0)\;,
\end{equation}
with initial condition
$G(\vec{x},t=0|\vec{x_0}) = \delta^d(\vec x - \vec x_0)$, and is given by
\begin{equation}
G(\vec x,t|\vec x_0) = \frac{ 1}{\left(4 \pi D t\right)^{d/2}} \exp
\left[-\frac{|\vec x-\vec x_0|^2}{4 D t}\right]\;.
\label{Gdef}
\end{equation}
Then the probability $p(\vec x,t)$ is a sum over two
contributions:  one which comes from trajectories where no resetting events
have occurred in time $t$ and a second contribution which comes from  summing
over trajectories where the last resetting event occurred at time $t-\tau$.
The probability of no resetting events having occurred up to time $t$  is ${\rm e}^{-rt}$
and the probability of the last resetting event having occurred
at $t-\tau$ (and no resetting events since) is $r {\rm e}^{-r\tau}$.
Thus the full time-dependent solution can be written down 
as 
\begin{equation}
p(\vec x,t) = {\rm e}^{-rt}  G(\vec x,t|\vec x_0) + r \int_0^t \D \tau \,
{\rm e}^{-r\tau}  G(\vec x,\tau|\vec X_r)\;.
\label{pt}
\end{equation}
The stationary state is attained when $t \gg 1/r$ where (\ref{pt}) tends to
the stationary distribution
\begin{eqnarray}
p^*(\vec x) =  r \int_0^\infty \D \tau\,  {\rm e}^{-r\tau }  G(\vec x,\tau|\vec X_r)\;.
\label{pss}
\end{eqnarray}
The relaxation to the steady state may be obtained from
\begin{equation}
p(\vec x,t) = p^*(\vec x) + {\rm e}^{-rt } G(\vec x,t|\vec x_0) - r \int_t^\infty \D \tau\, {\rm e}^{-r\tau } G(\vec x,\tau|\vec X_r)\;.
\end{equation}

In order to evaluate the integral in (\ref{pss})
we  use the identity (Equation 3.471.9 of \cite{GR})
\begin{equation}
\int_0^\infty \D t\, t^{\nu-1} {\rm e}^{-\frac{\beta}{t} -\gamma t}
= 2\left( \frac{\beta}{\gamma}\right)^{\nu/2}
K_\nu(2 \sqrt{\beta \gamma})\;,
\label{magint}
\end{equation}
where $K_\nu$ is the  modified Bessel function of the second kind (also known as Macdonald function) of order $\nu$. The relevant case of this identity
is 
\begin{equation}
\nu=1-d/2\;,
\end{equation}
and  one obtains from (\ref{Gdef}) and (\ref{pss}), after setting $\vec x_0 = \vec X_r$
\begin{equation}
p^*(\vec x) = \left( \frac{\alpha_0^2}{2\pi}\right)^{1-\nu} 
(\alpha_0 |\vec x- \vec X_r| )^\nu 
K_\nu(\alpha_0|\vec x- \vec X_r| )\;,
\label{pssgen}
\end{equation}
where
\begin{equation}
\alpha_0 =  \left( \frac{r}{D}\right)^{1/2}\;.
\label{alpha0}
\end{equation}
Expression (\ref{pssgen}), for the stationary distribution of  diffusion
in the presence of resetting in arbitrary dimension, is the central result of this section.
Note that (\ref{pssgen}) tends to zero as $|\vec x|\to \pm \infty$ and has a cusp at $\vec x= \vec X_r$ (see Figure 2).
Also note that  (\ref{pss1d}) is a nonequilibrium stationary state 
by which it is meant that there is circulation of probability
(even in a one-dimensional geometry). This is because resetting  
implies a source of probability at $\vec X_r$ while probability 
is lost from all other values of $\vec x \neq \vec X_r$.
\begin{figure}[ht]
  \begin{center}
    \includegraphics[width=0.8\textwidth]{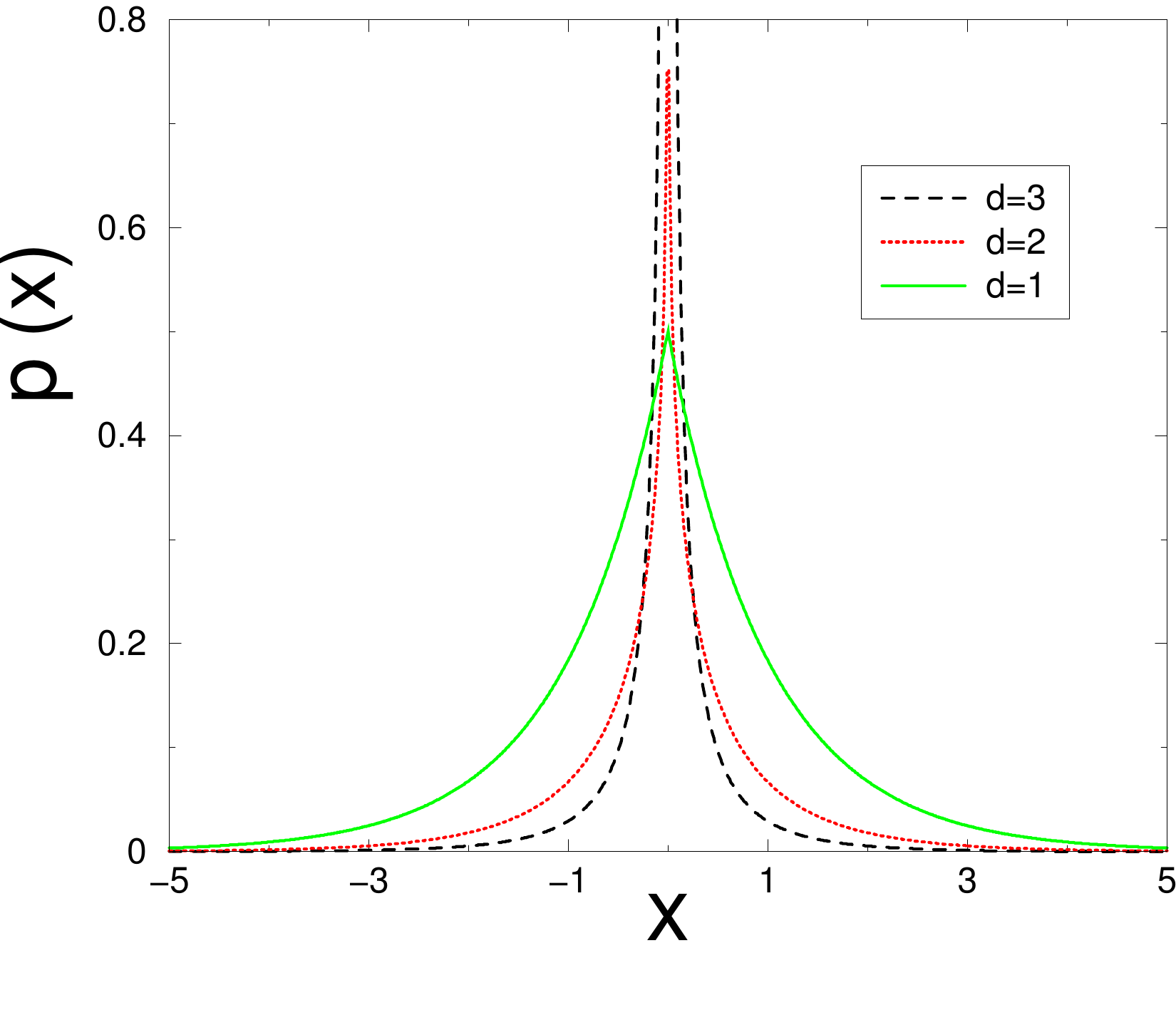}
  \end{center}
  \caption{The stationary probability densities $p^*(x= |\vec x|)$ given by Equation
\ref{pssgen} for the case $\alpha_0 =1$ and $\vec X_r=0$:
$d=1$ full lines; $d=2$ dotted lines; $d=3$ dashed lines.
}
    \label{fig2}
  \end{figure}

\subsubsection{Special Cases $d=1$ and $d=3$:}
In the case  $d=1$ ($\nu=1/2$) 
we use the identity
\begin{equation}
K_{1/2} (y) = \left( \frac{\pi}{2y}  \right)^{1/2} {\rm e}^{-y}
\label{Khalf}
\end{equation}
to find that  the  stationary distribution (\ref{pssgen}) reduces to 
\begin{equation}
p^*(x) = \frac{\alpha_0}{2} \exp( - \alpha_0 |x-X_r|)\;,
\label{pss1d}
\end{equation}
thus recovering the result of \cite{EM1},\cite{EM2}. 

Also in the case $d=3$ ($\nu=-1/2$), we may use the identity
$K_{-1/2} (y)=K_{1/2} (y)$ to find a simple form for the 
stationary distribution 
\begin{equation}
p^*(\vec x) = \frac{\alpha^2_0}{4 \pi |\vec x-\vec X_r|} \exp( - \alpha_0 |\vec x-\vec X_r|)\;.
\label{pss3d}
\end{equation}

\section{Survival probability in the presence of a trap at the origin}
\label{sec:sphd}

\subsection{Trap at origin and length scales of the system}
We now consider the presence of a trap at the origin, i.e.,
an absorbing  $d$-dimensional sphere of radius $a$
centred at  $\vec x=0$ which absorbs the particle.
The particle starts at the 
initial position $|{\vec x_0
}|>a$ and undergoes diffusion with diffusion constant $D$ and stochastic resetting
to  ${\vec X_r}$ with a constant rate $r$. When it reaches
the surface of the target sphere, the particle is absorbed  (see Figure 3).
\begin{figure}[ht]
  \begin{center}
    \includegraphics[width=0.8\textwidth]{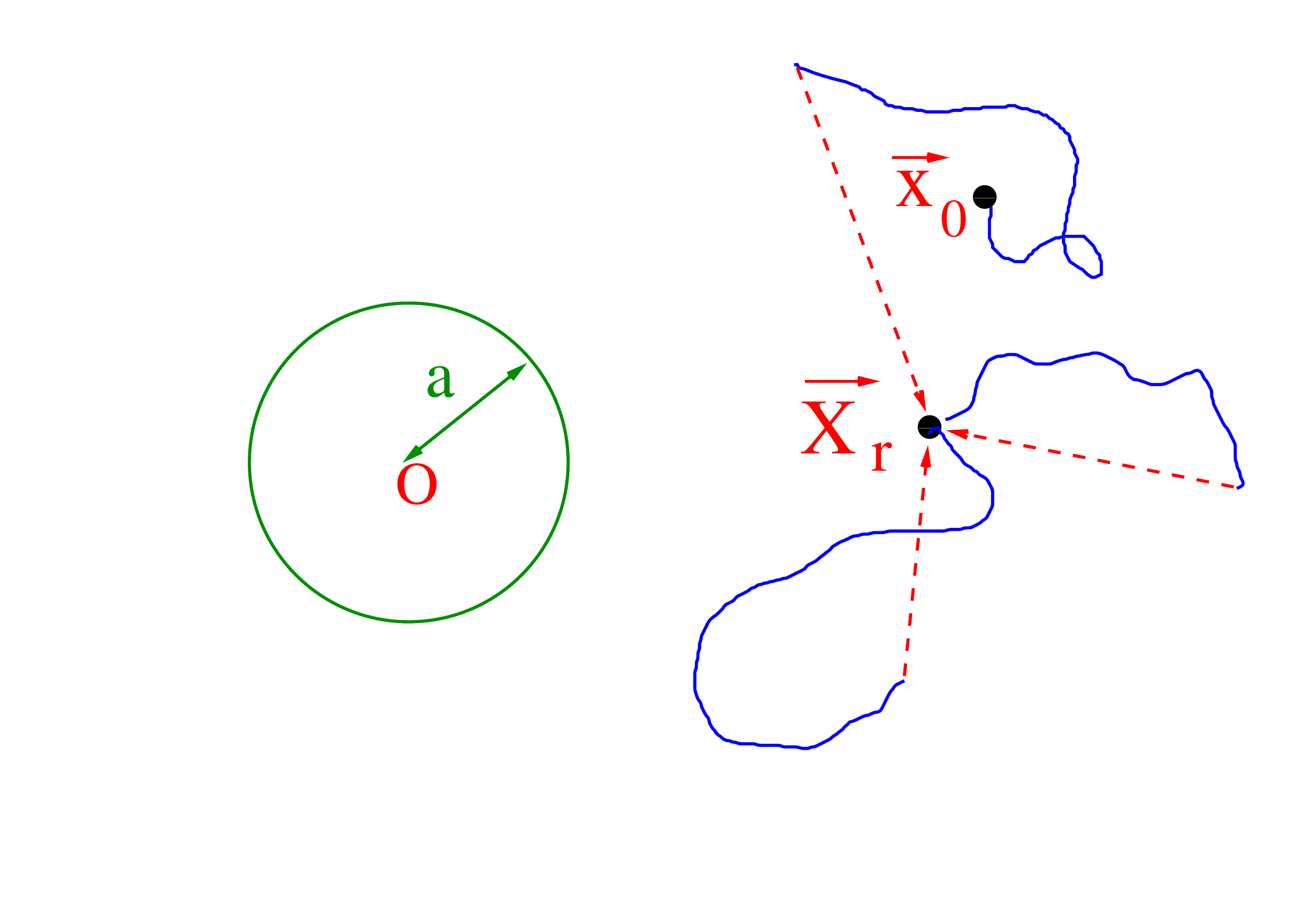}
  \end{center}
  \caption{Illustration in $d=2$ of  the diffusion of a particle with initial
position $\vec x_0$ and  resetting 
to $\vec X_r$, in the presence of an absorbing trap of radius $a$ with centre at the origin $O$.}
   \label{fig3}
  \end{figure}

Once we set $\vec x_0 = \vec X_r$, there are only three length scales
in the system.
The parameter $\alpha_0$ defined in (\ref{alpha0}) is an inverse length scale
which corresponds to the typical distance diffused by the particle
between resets. The other two length scales  
are  $R_r = |\vec X_r|$, the distance from the origin to resetting position, 
and $a$ the radius of the trap.
These three length scales can be combined
to define  dimensionless reduced variables
\begin{eqnarray}
\gamma &=& \alpha_0  R_r \label{gammadef} \\
\epsilon &=& \frac{a}{R_r}\label{epsdef}\;.
\end{eqnarray}
The dimensionless quantity $\gamma$ measures the ratio of the distance $R_r$ of the reset point  from the target at the origin to the 
diffusion length $\alpha_0^{-1}$;
$\epsilon$ is simply the ratio of the radius  of the absorbing sphere to 
 the distance $R_r$ of the reset point to the target at the  origin.

\subsection{Computation of survival probability}
To study absorption problems it is most convenient to consider the backward
master equation for the survival probability \cite{BMS13}.
In this formulation the initial position is considered as a variable.
In the presence of resetting it then is important to distinguish
the initial position which   is a variable (denoted henceforth by $\vec x_0$) from the resetting position $\vec X_r$ which is fixed. At the end of the calculation we will  set $\vec x_0 = \vec X_r$.

We are interested in the survival probability of a diffusive particle
at time $t$, having {\em started} at $\vec x_0$ at $t=0$ with resetting to $\vec X_r$, which in principle should be denoted $Q(\vec x_0,t; \vec X_r)$.
However, to lighten the notation we will suppress the resetting position
and write $Q(\vec x_0,t)$. In the same way we will write its Laplace transform,
defined below (\ref{qlthd}), as $q(\vec x_0,s)$.

The backward master equation for
$Q(\vec x_0,t)$
is constructed by considering events in the first infinitesimal time interval $\D t$: the components $(x_0)_i$ of the initial position become
\begin{eqnarray}
 (x_0)_i &\to& (X_r)_i\quad\mbox{with probability}\quad r \D t\nonumber\\
      &\to& (x_0)_i +\xi_i(0) \D t \quad\mbox{with probability}\quad 1-r \D t
\label{bfp}
\end{eqnarray}
where $\xi_i(0)$ is the $i$th component of the initial noise.
Then
\begin{equation}
Q(\vec x_0, t+\D t) = (1-r \D t) \langle Q(\vec x_0+\vec \eta(0) \D t, t) \rangle
+ r\D t Q(\vec X_r, t)\;,
\end{equation}
where $\langle\cdot\rangle$ indicates averaging over the initial noise.
Performing the average and taking the $\D t \to 0$ limit yields, for
 $|\vec x_0|>a$,
\begin{equation}
\frac{\partial Q(\vec x_0,t)}{\partial t}
= D \nabla^2_{\vec x_0} Q( \vec x_0, t)
 - r Q( \vec x_0, t)
 + r Q( \vec X_r, t)\;,
\label{bmehd}
\end{equation}
with boundary and initial conditions $Q(|\vec x_0|=a,t)=0$ and $Q(\vec x_0, 0)=1$ for $|\vec x_0|>a$. 

The Laplace transform
\begin{equation}
q(\vec x_0, s) = \int_0^\infty\D t\ {\rm e}^{-st} Q(\vec  x_0, t)
\label{qlthd}
\end{equation}
satisfies
\begin{equation}
 D \nabla^2_{\vec x_0}\, q(\vec{x_0}, s)
 - (r+s) q(\vec x_0, s)   = -1 - r  q(\vec X_r, s ) \;.
\label{hdbeq}
\end{equation}
The solution to the homogeneous equation
\begin{equation}
 D \nabla^2_{\vec x_0}\, q(\vec x_0, s)
 - (r+s)  q(\vec x_0, s)  = 0\;,
\end{equation}
which is radially symmetric about the origin  and does not diverge as $|\vec x_0|\to \infty$, is
\begin{equation}
 q_{\rm hom}(\vec x_0, s) =
(\alpha  |\vec x_0|)^\nu K_\nu( \alpha |\vec x_0|)
\end{equation} 
where now
\begin{equation}
\alpha(s) = \left( \frac{r+s}{D}\right)^{1/2}\;,
\end{equation}
and as before, 
$\nu = 1- d/2$. Note that $\alpha(0) = \alpha_0$  given in (\ref{alpha0}).
Hereafter, we write $\alpha \equiv \alpha(s)$ for brevity.

In order to determine the solution to (\ref{hdbeq})
we write
\begin{equation}
q(\vec{x_0}, s) =
B  |\vec x_0|^\nu K_\nu( \alpha |\vec x_0|) +C
\label{qhd2}
\end{equation}
where $C$ is independent of $|\vec x_0|$. Then the absorbing boundary condition implies that $C = -B  a^\nu K_\nu( \alpha a)$ and substituting (\ref{qhd2}) back into 
(\ref{hdbeq})  determines
$$
B= - \frac{1}{  s a^\nu K_\nu( \alpha a) + r |\vec X_r |^\nu K_\nu( \alpha |\vec X_r| ) }\;.
$$
Therefore
\begin{equation}
q(\vec x_0,s )
= 
\frac{
 a^\nu  K_\nu( \alpha a)- |\vec x_0 |^\nu 
 K_\nu( \alpha |\vec x_0 |)
 }
{r |\vec X_r |^\nu 
 K_\nu( \alpha |\vec X_r|)+ sa^\nu 
 K_\nu( \alpha a)
}\;.
\label{qhdfin}
\end{equation}
In particular, setting the starting position $\vec x_0$ equal to the resetting position $\vec X_r$, and using isotropy,  we have 
\begin{equation}
q(\vec X_r,s )  =q(R_r,s )  
= 
\frac{
 a^\nu  K_\nu( \alpha a) - R_r ^\nu 
 K_\nu( \alpha R_r)
 }
{r  R_r^\nu 
 K_\nu( \alpha R_r)+ sa^\nu 
 K_\nu( \alpha a)
}\;,
\label{qhdfin}
\end{equation}
where  $R_r =|\vec X_r |$.
Equation (\ref{qhdfin}) is the main result of this section
and is an exact expression for the Laplace transform of the survival probability in arbitrary dimension $\nu=1- d/2$.

Having obtained   this  expression one would ideally wish to invert the Laplace transform in order to obtain 
\begin{equation}
Q(R_r,t)\equiv Q(|\vec x_0|= R_r,t; |\vec X_r|= R_r)\;.
\end{equation}
However, it appears difficult to invert the
Laplace transform (\ref{qhdfin}) explicitly for all parameters. 
Nevertheless there are two limits where one may extract the forms of
$Q(R_r,t)$:
(i) the  natural scaling regime $r \to 0$, $t\to \infty$ with scaling variable $y=rt$ fixed; (ii) for fixed $r$, the large $t$ limit.

\subsection{Scaling regime}
In the scaling regime,
we consider  the  limit $r \to 0$, $s\to 0$ where
\begin{equation}
\lambda = \frac{s}{r}\;.
\label{lambdef}
\end{equation}
The scaling variable is
\begin{equation}
y  = rt\;.
\end{equation}
In this limit we  obtain the scaling distribution
\begin{equation}
Q(R_r,t) \to  F(R_r,y)
\end{equation}
and (\ref{qhdfin}) becomes
\begin{equation}
q(R_r,s) \to f(R_r,\lambda)  =\int_0^\infty\D y\, {\rm e}^{-\lambda y} \frac{F(R_r,y)}{r}
\label{ltsl}
\end{equation}
i.e. we obtain a simple  scaling form of the Laplace transform 
$f(R_r,\lambda)$. We may then  invert  the Laplace transform in (\ref{ltsl}) with respect to $\lambda$ to obtain the scaling form
$F(R_r,y)$ for the survival probability.

To proceed,  we  use the small argument expansion
\begin{eqnarray}
 K_\nu(x) &\simeq&  - \ln (x/2) - \gamma_E \quad  \mbox{if }\quad \nu=0
\label{nu0} \\[1ex] 
&\simeq& \frac{\Gamma(\nu)}{2} \left( \frac{2}{x} \right) ^{\nu}
+ \frac{\Gamma(-\nu)}{2} \left( \frac{2}{x} \right) ^{-\nu} + \ldots
\quad   \mbox{if }\quad 0< \nu <1, 
\label{nupos}
 \\ 
&\simeq& \frac{\Gamma(\nu)}{2} \left( \frac{2}{x} \right) ^{\nu}
- \frac{\Gamma(\nu-1)}{2} \left( \frac{2}{x} \right) ^{\nu-2} + \ldots
\quad   \mbox{if }\quad \nu > 1,  
\label{nug1}
\end{eqnarray}
where $\gamma_E = 0.5772\ldots$ is the Euler constant.\\[1ex]

We consider separately the different cases $\nu >0$, $\nu =0$,
$\nu <0$ corresponding to $d<2$, $d=2$, $d>2$ respectively
(using $\nu=1-d/2$).\\[1ex]

\noindent {\bf i)  $1>\nu >0$ $(0<d<2)$}:
In this case both terms in (\ref{nupos}) are required and one finds
\begin{equation}
q(R_r,s) \simeq  \frac{\Gamma(-\nu)}{\Gamma(\nu)}
\frac{r^{\nu-1}(1+\lambda)^{\nu-1}}{(4D)^\nu}\left[ a^{2\nu}-R_r^{2\nu}\right]\;.
\label{qlsld1}
\end{equation}
One may then invert (\ref{qlsld1}) using
(\ref{ltsl}) to obtain the final result in the scaling limit
\begin{equation}
Q(R_r,t)= F(R_r,rt)
\end{equation}
where the scaling function is given by
\begin{equation}
F(R_r,y) = \frac{1}{\nu \Gamma(\nu)} \left( \frac{\gamma}{2}\right)^{2\nu}
(1- \epsilon^{2\nu}) {\rm e}^{-y}y^{-\nu}\;,
\end{equation}
and,  as usual, $\gamma = R_r(r/D)^{1/2}$ and $\epsilon =a/R_r$.

In particular, the $d=1$ ($\nu=1/2$) case is
\begin{equation}
F(R_r,y) = 
\frac{\gamma}{\sqrt{\pi}}
\frac{{\rm e}^{-y}}{y^{1/2}}\;.
\end{equation}
Note that in the limit $y\to 0$
\begin{equation}
 F(R_r,y) \simeq
\frac{R_r}{\sqrt{D \pi}}\left(\frac{r}{y}\right)^{1/2} =
\frac{R_r}{\sqrt{D \pi t}}\;,
\end{equation}
which recovers the usual 1$d$ diffusive survival probability at large $t$.
Also note that for $y$ large the survival probability decays exponentially.
\vspace*{2ex}

\noindent {\bf ii) $\nu =0\; (d=2)$}:
In the case $d=2$, which corresponds to $\nu=0$, the leading order expansion 
of (\ref{qhdfin}) in the limit $r \to 0$, $s \to 0$ is
\begin{equation}
q(R_r,s) \simeq \frac{2\ln \epsilon}{(s+ r)\ln r}\;.
\end{equation}
Inverting the Laplace transform yields
\begin{equation}
Q(R_r, t) \simeq \frac{2 \ln \epsilon}{\ln r}\,  {\rm e}^{-rt}\;.
\end{equation}
So in the case $d=2$  a scaling form, which is solely a function of $y$ and $\epsilon$, does not emerge.\\[1ex]

\noindent {\bf iii) $\nu <0\; (d>2)$.}
In the case $d>2$, which corresponds to $\nu<0$,
we can use the symmetry   $K_{-\nu}(x) = K_\nu(x)$.
Then it turns out that only the leading order term in the small $x$ expansions
(\ref{nupos},\ref{nug1})
matters. Substituting this in
 (\ref{qhdfin}) gives  in the scaling  limit  
\begin{equation}
q(R_r,s) \simeq \frac{1}{r}\frac{1-\epsilon^{-2\nu}}{(\lambda + \epsilon^{-2\nu})}\;.
\end{equation}
Inverting the Laplace transform yields
\begin{equation}
F(R_r,y)= (1-\epsilon^{-2\nu}) {\rm e}^{-y\epsilon^{-2\nu}}\;.
\end{equation}
Thus for $d>2$ the  scaling form decays exponentially in the scaling variable $y$.

\subsection{Long-time asymptotics of Survival Probability}

The long-time asymptotics of the survival probability for fixed $r$ may be deduced from the inversion of
(\ref{qhdfin})
\begin{equation}
 Q(R_r,t) =\int_C \frac{\D s\, {\rm e}^{st}}{2 \pi i} 
\frac{
 a^\nu  K_\nu( \alpha a) -  R_r^\nu 
 K_\nu( \alpha R_r) }
{r R_r^\nu 
 K_\nu( \alpha R_r)+ sa^\nu 
 K_\nu( \alpha a)
}\;,
\label{Qinv}
\end{equation}
where $C$ is the Bromwich contour to the right of any singularities in the complex $s$ plane.
We may rewrite (\ref{Qinv}) as
\begin{equation}Q(R_r,t) =\int_C \frac{\D s}{2 \pi i} \frac{{\rm e}^{st}}{r}
\left[ \frac{r+ g(s)}{s-g(s)}\right]
\label{Qinv2}
\end{equation}
where
\begin{equation}
g(s) =  - \frac{r}{\epsilon^\nu} \frac{K_\nu(\alpha(s) R_r)}{ 
K_\nu(\alpha(s) a)}\;,
\end{equation}
with $\epsilon = a/R_r$ and $\displaystyle \alpha(s) = \sqrt{\frac{r+s}{D}}$.
The analytic structure in the complex $s$ plane
of the integrand of (\ref{Qinv2})
is a branch point $s = -r$ coming from $\alpha(s)$, and a simple pole at $s=s_0$ which satisfies
\begin{equation}\label{eq:pole}
s_0 = g(s_0) \;.
\end{equation}
In the long-time limit the dominant contribution to
(\ref{Qinv2}) will come from the pole at $s_0$ which yields by the residue theorem
\begin{equation}
Q(R_r,t) \simeq  \frac{{\rm e}^{s_0t}}{r}\frac{[r+s_0]}{1-g'(s_0)}\;.
\label{Qas}
\end{equation}

Making the substitution
\begin{equation}
s_0 = r(u_0-1)\qquad 0 < u_0 <1\;,
\label{sdef}
\end{equation}
equation (\ref{eq:pole}) becomes a transcendental equation for $u_0$
\begin{equation}
u_0 = 1 - \frac{1}{\epsilon^\nu} 
\frac{K_\nu(\gamma u_0^{1/2})}{K_\nu(\gamma \epsilon  u_0^{1/2})}\;,
\label{u0eqn}
\end{equation}
where $\gamma = \alpha_0 R_r$.
Thus the asymptotic behaviour of the survival probability is
\begin{equation}
\ln Q(R_r,t) \sim r(u_0-1)t
\end{equation}
where $u_0$ satisfies (\ref{u0eqn}).

In the limit $\gamma \gg 1$ (and $\gamma \epsilon \gg 1$)
we can use the asymptotic behaviour
\begin{equation}
K_{\nu}(y)\simeq \left( \frac{\pi}{2y}\right)^{1/2} \, {\rm e}^{-y}\left[1 + O(1/y)\right] \quad \mbox{for large}\quad y
\label{Kasym}
\end{equation}
to obtain the  behaviour of $u_0$
 \begin{equation}
u_0 \simeq  1 - \epsilon^{1/2-\nu} 
{\rm e}^{-\gamma(1-\epsilon)}\;.
\label{u0asy}
\end{equation}
Then (\ref{Qas}) takes the form
\begin{equation}
Q(R_r,t) \sim  \exp \left(
- rt \epsilon^{1/2-\nu} 
{\rm e}^{-\gamma(1-\epsilon)}\right)\;.
\label{Qasymd}
\end{equation}
This expression has the form of a Gumbel distribution, the origin
of which we shall now discuss.

\subsubsection{Renewal process and extreme value statistics interpretation}
\label{renew}
In order to better understand the  result (\ref{Qasymd}) we  consider the
diffusion with resetting process as a  process which is renewed each time
the particle is reset---the renewed process has no memory of  its past history.
We shall consider the long time regime where the number of resets
to position $\vec X_r$ in  time $t$ is the mean number of resets
$N=rt$ plus fluctuations of $O(t^{1/2})$.

In order for the target to survive up to time $t$ it must survive
through each of the resets.
Then we can write the asymptotic behaviour as
\begin{equation}
Q(R_r,t)  \sim  \left(\langle P^{\rm diff}_S(\tau)\rangle_{\tau} \right)^{rt}\;,
\end{equation}
where the $P^{\rm diff}_S$ is the survival probability of a diffusive particle
up to time $\tau$ and $\langle \cdot \rangle_{\tau}$ denotes an average over  the 
duration  of the reset (time until next resetting event). As resetting is a Poisson process we have
\begin{equation}
\langle P^{\rm diff}_S(\tau)\rangle_{\tau}
=\int_0^\infty \D \tau\, 
r e^{-r\tau} P^{\rm diff}_S(\tau) = r q^{\rm diff}(R_r, r)\;,
\end{equation}
where now $q^{\rm diff}(R_r, r)$ is the Laplace transform, with Laplace
variable $s$ replaced by $r$, of the survival
probability of a diffusive particle with an absorbing target at the origin.
The expression for $q^{\rm diff}(R_r, r)$ is easy to construct using, for example, a backward
master equation approach and is well-known in the literature (see e.g. \cite{Redner}): 
\begin{eqnarray}
q^{\rm diff}(R_r, r)
&=&\frac{1}{r}\left[ 1-  
\left( \frac{ R_r}{a}\right)^\nu 
\frac{ K_\nu(  R_r (r/D)^{1/2})}{K_\nu(a (r/D)^{1/2})}
\right]\\
&=&\frac{1}{r}\left[ 1-  
 \epsilon^{-\nu} 
\frac{ K_\nu(  \gamma )}{K_\nu(\gamma \epsilon)}
\right]\;.
\end{eqnarray}
We then have
\begin{eqnarray}
Q(R_r,t)  &\sim& \exp\left\{  rt
\ln \left[ 1-  
 \epsilon^{-\nu} 
\frac{ K_\nu(  \gamma )}{K_\nu(\gamma \epsilon)}
\right]\right\}\\
&\simeq& \exp \left[ -rt
 \epsilon^{1/2-\nu} 
   {\rm e} ^{-\gamma(1-\epsilon)}\right]\;,
\end{eqnarray}
where we have used  the asymptotic behaviour (\ref{Kasym})
for large $\gamma$ and $\epsilon \gamma$, thus recovering (\ref{Qasymd}).

One may relate this result to the Gumbel distribution
for the extremum of independent random variables~\cite{Gumbel}
as follows. As we have seen above, for the target to survive
it must survive through $N$ resets. Thus the leftmost point
the searcher has reached in the $N$ resets must be less than
$R_r$, the distance to the target.
In other words, the maximum of $N$ independent random variables must be less than $R_r$. Thus the Gumbel distribution, which is the scaling form for the cumulative distribution of the
maximum of $N$ independent random variables with distribution decaying faster than a power law, naturally emerges.

\subsection{Generalisation to a resetting distribution}
The computation of the survival probability can easily be generalised to the case of a resetting distribution
${\cal P}(\vec z)$ \cite{EM2}
where  $\vec z$ is a randomly chosen resetting position.
Equation (\ref{rule.1}) is modified to
\begin{eqnarray}
 x_i(t+dt) & =&  z_i \quad {\rm with\,\,probability}\,\, r {\cal P}(\vec z) \, dt \nonumber \\
& =&  x_i(t) + \xi_i(t) dt \quad {\rm with\,\,probability}\,\, (1-r\, dt)\;.
\label{rule.1a}
\end{eqnarray}
Also, if we  choose the distance of the initial position from the origin, $R= |\vec x_0|$ to be drawn from the same distribution
${\cal P}(R)$
we find that the Laplace transform of the survival probability
averaged over the resetting distribution
becomes 
\begin{equation}
\overline{q}(s) 
= \frac{a^\nu 
 K_\nu( \alpha a)
- \int \D R\, R^\nu 
 K_\nu( \alpha R) {\cal P}(R)}
{sa^\nu 
 K_\nu( \alpha a)
+r 
 \int \D R\, R^\nu 
 K_\nu( \alpha R) {\cal P}(R)
}\;.
\label{qhdgen}
\end{equation}
Equation (\ref{qhdgen})  gives the survival probability for a diffusive particle
with a trap at the origin for abitrary resetting distribution,
thus generalising the result  
(\ref{qhdfin}) for a fixed resetting position.
Equation (\ref{qhdgen}) may be used to study the optimal choices of resetting distribution  to optimize for example the mean time to absorption \cite{EM2}.

\section{Mean First-passage time to absorption by sphere of radius $a$ in arbitrary dimensions}

We now consider the mean time to absorption in the presence of an absorbing  trap in the case of diffusion with resetting to point $\vec X_r$.

\subsection{Mean time to absorption}

The mean time to absorption  by the trap at the origin, with coincident initial and resetting positions,
can be computed from the survival probability $Q(R_r,t)$
\begin{eqnarray}
T(R_r)  &=& - \int_0^\infty \D t\, t \frac{\partial Q(R_r,t)}{\partial t}\nonumber\\
&=& q(R_r,s=0)\;.
\end{eqnarray}
(The integration by parts requires $Q(R_r,t)$ to vanish faster than $1/t$ for large $t$.)
Thus, the mean time to absorption 
is obtained by setting $s=0$ in (\ref{qhdfin})
\begin{equation}
T(R_r) = \frac{ 1}{r}\left[
\left( \frac{a}{R_r}\right)^\nu 
\frac{K_\nu( \alpha_0 a)}
{ K_\nu( \alpha_0 R_r)} -1 \right]\;.
\label{Td}
\end{equation}
Note that for $d\leq 2$, (\ref{Td})
has a finite limit as $a \to 0$,
but for $d > 2$, (\ref{Td})
diverges as $a \to 0$.

Let us recall that
\begin{equation}
K_{\nu}(z)= K_{-\nu}(z)\;.
\label{mbfp.1}
\end{equation}
Then using the dimensionless quantities
 (\ref{gammadef}), (\ref{epsdef}) one may rewrite  (\ref{Td}) as
\begin{equation}
T = \frac{1}{\gamma^2}\,\left[\epsilon^{\nu}\, 
\frac{K_{|\nu|}(\epsilon\,\gamma)}{K_{|\nu|}(\gamma)}-1\right]\, \frac{R_r^2}{D}\,,
\label{mfpt.2}
\end{equation}
where, as usual,  $\gamma= \alpha_0 R_r = \sqrt{r/D} R_r$
and  we have suppressed the $R_r$ dependence of $T(R_r)$ for convenience.
The first point to note is that $T$ is finite for $0< r<\infty$. It diverges
as $r \to 0$ as $T \sim r^{-1/2}$ in $d=1,3$  and as $T \sim 1/(r |\ln r|)$ in $d=2$
as can be checked using the small argument expansion of the modified Bessel functions (\ref{nu0}--\ref{nug1}).
It also diverges exponentially in $r$ as $r \to \infty$
which can be easily checked   using the asymptotic behaviour  (\ref{Kasym}).
The $r \to 0$ limit  recovers the well-known result
that the mean time for a diffusive particle to reach the origin diverges.
The $r\to \infty$   limit
merely expresses the fact 
that as the reset rate increase the diffusing particle 
has less time between resets to  reach the origin.

Now, for fixed $\epsilon$ and fixed $R_r^2/D$, we wish to see how $T$ in Eq. 
(\ref{mfpt.2}) behaves as a function of the reduced variable $\gamma=\sqrt{r/D}\,R_r$.
Since  $T$ diverges in the limits
$\gamma\to 0$ and $\gamma\to \infty$, 
there is a unique minimum of $T$ at the optimal value $\gamma^*$ where $\D T/\D \gamma=0$.
This optimal $\gamma^*$ is evidently a complicated function of $\epsilon$ and in
general, is hard to determine explicitly. However, we are interested in the limit
$\epsilon=a/R_r\to 0$ limit such that the target size is small compared to
the particles initial position. In this limit, one can obtain explicit results as we now show.

\subsection{Small $\epsilon$ limit}
Taking $\epsilon\to 0$ limit in Eq. (\ref{mfpt.2}), upon using the asymptotic behaviour
of $K_{\nu}(z)$ in Eqs. (\ref{nu0}) and (\ref{nupos}), gives for $\nu=1-d/2\ne 0$
\begin{equation}
T\to \frac{1}{\gamma^2}\, \left[\epsilon^{\nu-|\nu|}\, \Gamma(|\nu|)\, 2^{|\nu|-1}\, 
\frac{\gamma^{-|\nu|}}{K_{|\nu|}(\gamma)}-1\right]\, \frac{R_r^2}{D}\, .
\label{eps0.1}
\end{equation}

\subsubsection{Case $d<2$:}
Consider first the case $d<2$. In this case, $\nu=1-d/2$ and $|\nu|=1-d/2$ and hence
the $\epsilon$ dependence of $T$ in Eq. (\ref{eps0.1}) just drops out in the $\epsilon\to 
0$ limit. This is consistent with the fact that for $d<2$, one does not need to have 
a finite size of the target and there is well defined point target limit. Hence, for $d<2$, 
we find the exact result
\begin{equation}
T_{d<2} = \frac{1}{\gamma^2}\,\left[2^{-d/2}\, \Gamma\left(1-\frac{d}{2}\right)\, 
\frac{\gamma^{d/2-1}}{K_{1-\frac{d}{2}}(\gamma)}-1\right]\, \, \frac{R_r^2}{D}.
\label{dlt2.1}
\end{equation}
In particular, for $d=1$,
using  (\ref{Khalf}) we recover the $1$-d result \cite{EM1}
\begin{equation}
T_{d=1}= \frac{1}{\gamma^2}\left[{\rm e}^{\gamma}-1\right]\, \frac{R_r^2}{D},
\label{1d.1}
\end{equation}
which has a unique minimum at $\gamma^*=\gamma_1= 1.59362\dots$ where $\gamma_1$ is the 
unique positive root of $\gamma_1-2(1-e^{-\gamma_1})=0$.
For general $d<2$, one can similarly minimize Eq. (\ref{dlt2.1}) and obtain
the optimal $\gamma^*$.

\subsubsection{Case $d>2$:}
In the opposite case $d>2$, the $\epsilon$ dependence of $T$ does not drop out
of Eq. (\ref{eps0.1}), indicating that for $d>2$, one needs a finite target size
as otherwise a point particle will never meet a point target. In this case, to leading 
order in $\epsilon$, Eq. (\ref{eps0.1}) gives
\begin{equation}
T_{d>2}\to \left[\frac{\gamma^{-1-d/2}}{K_{\frac{d}{2}-1}(\gamma)}\right]\, 
\Gamma\left(\frac{d}{2}-1\right)\, 
2^{d/2-2}\, \epsilon^{2-d}\,\frac{R_r^2}{D}\, .
\label{dgt2.1}
\end{equation}
One can then minimize $T$ as a function of $\gamma$. The optimal value $\gamma^*$ 
is obtained by setting $\displaystyle \frac{\D T}{\D \gamma}=0$ and $\gamma=\gamma^*$. Taking
the derivative
of Eq. (\ref{dgt2.1}) with respect to $\gamma$ and using the property of the modified Bessel 
function \cite{GR}, 
\begin{equation}
y K_{\nu}'(y)+ \nu K_{\nu}(y)=-y K_{\nu-1}(y),
\end{equation}
we obtain $\gamma^*$
as a root of the equation
\begin{equation}
2 K_{\frac{d}{2}-1}(\gamma^*)-\gamma^*\, K_{\frac{d}{2}-2}(\gamma^*)=0\,.
\label{dgt2.2}
\end{equation}
A particular simplification occurs for $d=3$ where one can use 
$K_{-1/2}(y)= K_{1/2}(y)$ to obtain explicitly $\gamma^*=2$.
Hence, using $K_{1/2}(y)=\sqrt{\pi/{2y}}\, e^{-y}$, the optimal mean first-passage time for $d=3$ is given explicitly by
\begin{equation}
T^*_{d=3}= {\rm e}^2\, \frac{R_r^2}{D}\, \frac{1}{\epsilon}\, .
\label{d3.1}
\end{equation}

In the marginal dimension $d=2$, one has $\nu=0$ and hence from Eq. (\ref{mfpt.2})
we obtain
\begin{equation}
T_{d=2}= \frac{1}{\gamma^2}\, \left[\frac{K_0(\epsilon\,\gamma)}{K_0(\gamma)}-1\right]\, 
\frac{R_r^2}{D}\,.
\label{d2.1}
\end{equation}
Taking again the $\epsilon\to 0$ limit using Eq. (\ref{nu0}) gives,
to leading order in small $\epsilon$,
\begin{equation}   
T_{d=2}\to \frac{-\ln(\epsilon)}{\gamma^2\, K_0(\gamma)}\, \frac{R_r^2}{D}\,.
\label{d2.2}
\end{equation}
Once again, one can minimize $T$ as a function of $\gamma$. The optimal
$\gamma^*$ is obtained by solving $\displaystyle \frac{\D T}{\D \gamma}=0$ and 
is given by the root of the equation 
\begin{equation} 
2K_{0}(\gamma^*)-\gamma^*\, K_{1}(\gamma^*)=0
\label{d2.3}
\end{equation}
which gives $\gamma^*(d=2)= 1.55265\dots$.
Substituting this result in Eq. (\ref{d2.2}), we find that the exact optimal mean first-passage time in $d=2$
\begin{equation}
T^*_{d=2}\to 2.07679 \dots\, [-\ln (\epsilon)]\, \frac{R_r^2}{D}\, .
\label{d2.4}
\end{equation}

\section{Many Searchers}

\subsection{Average and typical behaviour: annealed and quenched averages}

We now consider the problem of many independent searchers (diffusive
particles) and the survival probability of a stationary target at the
origin defined as in Section 4.1.  Specifically, we consider $N$
diffusive particles labelled $\mu =1,\ldots, N$, each of which is
reset to its own resetting position $\vec X_\mu $ with rate $r$. We
also take the initial position of each searcher to be identical to
its resetting position.

The survival probability of the target is given by
\begin{equation}
P_s(t) = \prod_{\mu=1}^N Q(\vec X_\mu,t)
\label{Psdef}
\end{equation}
where 
$Q(\vec X_\mu,t)$ is the survival probability in the single searcher problem
considered in Section \ref{sec:sphd}. Note that in the absence of resetting ($r=0$) this survival probability  $P_s(t)$ has been studied in all dimensions
\cite{Tachiya, Klafter,bb02,bb03}.

We  consider the $N$ resetting positions to be distributed uniformly with density $\rho$ outside of the target volume  and 
consequently, $P_s(t)$ is a random variable. 
Its  average is simply $P_s^{\rm av}(t)=\langle P_s(t)\rangle_{\vec x}$ where $\langle \cdot  \rangle_{\vec X}$ denotes averages over $\vec X_\mu$'s.
However, as we shall see,  for a {\em typical} resetting  configuration
$P_s(t)$ is not captured by the average. 
This is because the average may be dominated by rare distributions
of the resetting  positions of the searchers for which the survival probability is much larger  than is typical.
One can think of the resetting positions as initial conditions which are remembered for all time by the system through the resetting dynamics.

The typical $P_s(t)$ can be extracted
by first averaging over the logarithm of $P_s(t)$ followed by 
exponentiating: $P_s^{\rm typ}(t)= \exp\left[\langle \ln P_s(t) \rangle_{\vec X} 
\right]$. One can draw an analogy to a disordered system with $P_s(t)$
playing the role of partition function $Z$ and $X_\mu$'s corresponding to
disorder variables.
Thus the average and typical behaviour correspond respectively 
to the {\em annealed} average (where one averages the partition function $Z$)
and the {\em quenched} average (where one averages the free energy $\ln Z$)
in disordered systems.

To  compute the average behaviour of (\ref{Psdef}) (the annealed case)
we may write
\begin{eqnarray}
P_s^{\rm av}(t) &=&
 \left[ \langle  Q(\vec X,t)\rangle_{\vec X}\right]^N \\
&=&  \exp\left\{ N \ln\left[1- \langle 1- Q(\vec X,t) \rangle_{\vec X}\right] \right\}
\end{eqnarray}
where $\langle \cdot\rangle_{\vec X}$ denotes an average over the
resetting position $\vec X$. We begin by considering $\vec X$ to be
distributed uniformly over a volume $V$ comprising a sphere of
radius $L$ with the target volume, which is a sphere of radius $a$,
removed. Noting that $Q(\vec X,t)=Q(R,t)$, where $R= |\vec X|$, we obtain
\begin{equation}
\langle 1-  Q(\vec X,t)\rangle_{\vec X}
=
 1 - \frac{1}{V} \int_{R>a}\D R\, \Gamma_d R^{d-1} 
\left( 1- Q(R,t)\right) \;,
\end{equation}
where 
\begin{equation}
\Gamma_d =  \frac{2\pi^{d/2}}{\Gamma(d/2)}
\end{equation}
is the surface  area of a  $d$-dimensional unit sphere.
Letting $N,L \to \infty$
but keeping the density of walkers $\rho=N/{V}$ fixed, 
we obtain
\begin{equation}
P_s^{\rm av}(t) \to  \exp\left[ -\rho \Gamma_d  \int_a^\infty\D R\, R^{d-1}  \left(1-  Q(R,t)\right)\right]
\equiv\exp\left[ - \rho I_1(t)\right]\;.
\label{Psann}
\end{equation}

On the other hand, the typical behaviour (the quenched case) 
$P_s^{\rm typ}(t)  = \exp \left[\langle \ln \left( P_s(t)\right) \rangle_X\right]$
can be expressed as
\begin{equation}
P_s^{\rm typ}(t) =
   \exp \left\{  \sum_{\mu=1}^N  \langle \ln\left[ Q(\vec X_\mu,t)\right] \rangle_{\vec X_\mu} \right\}
= \exp \left\{ N \langle \ln\left[ Q(\vec X,t)\right] \rangle_{\vec X}\right\}
\end{equation}
where 
\begin{equation}
\langle \ln\left[ Q(\vec X,t)\right] \rangle_{\vec X} 
= \frac{1}{V} \int_{R>a}\D R\, \Gamma_d R^{d-1}  \ln\left[ Q(R,t)\right]\;.
\end{equation}
In the limit  $N,L \to \infty$
with  density of walkers $\rho=N/{V}$ fixed, 
we obtain
\begin{equation}
P_s^{\rm typ}(t) 
=  \exp \left[   \rho \Gamma_d \int_a^{\infty} \D R R^{d-1} \ln  Q(R,t)\right] 
\equiv\exp\left[ - \rho I_2(t)\right]\;.
\label{PsI2}
\end{equation}
Thus the determination of the average and typical behaviour reduces to the  evaluation of two integrals:
\begin{eqnarray}
I_1 &=& \Gamma_d \int_a^\infty\D R\, R^{d-1}  [1-  Q(R,t)]\label{I1def} \\
I_2 &=&  - \Gamma_d \int_a^\infty\D R\, R^{d-1}  \ln\left[ Q(R,t)\right]\label{I2def}\;.
\end{eqnarray}

\subsection{Average behaviour}
\label{sec:Psa}
The Laplace transform of (\ref{I1def}) ${\tilde I_1}(s)=\int_0^{\infty} I_1(t) e^{-st} \D t$
can be determined using (\ref{qhdfin}):
\begin{eqnarray}
{\tilde I_1}(s) &=&
\Gamma_d \int_a^\infty\D R\,R^{d-1}\left[\frac{1}{s}-q(R,s)\right]
\nonumber \\
&=& \Gamma_d \frac{(r+s)}{s} \int_a^\infty\D R\,R^{d-1}\left[r + s \left( \frac{a}{R}\right)^\nu 
\frac{K_\nu(\alpha a)}{K_\nu(\alpha R)}\right]^{-1}\;.
\label{int11}
\end{eqnarray}
We now wish to  determine the small $s$ behaviour. First note that for large $x$
\begin{equation}
K_\nu(x)\sim \left(\frac{\pi}{2x}\right)^{1/2} {\rm e}^{-x}\;.
\label{mbfxgg1}
\end{equation}
Therefore in the integral (\ref{int11}) the term $r$ in the square bracket will dominate upto some
length scale $R^*$ and we can determine the leading behaviour as
\begin{eqnarray}
{\tilde I_1}(s) &\simeq&
\Gamma_d \frac{(r+s)}{rs} \int_a^{R^*} \D R\,R^{d-1} \\
&\simeq &
\Gamma_d \frac{(r+s)}{rs} \frac{(R^*)^{d}}{d} \;,
\end{eqnarray}
where we have assumed $R^* \gg a$.

In order to deduce $R^*$ we note that it is defined by
\begin{equation}
 \left( \frac{a}{R_r}\right)^\nu 
\frac{K_\nu(\alpha a)}{K_\nu(\alpha R_r)} \simeq \frac{r}{s}\;.
\end{equation}
Using  the asymptotic behaviour
(\ref{mbfxgg1}) of $K_\nu(\alpha R^*)$ for $R^*$ large we  have
\begin{equation}
\frac{{\rm e}^{\alpha_0R^*}}{(R^*)^{\nu-1/2}} \simeq 
\left(\frac{\pi}{2\alpha_0}\right)^{1/2}\frac{1}{a^\nu K_\nu(\alpha_0 a)}\frac{r}{s}\;,
\end{equation}
therefore as $s\to 0$ we obtain
\begin{equation}
R^* \simeq \frac{1}{\alpha_0} \ln \left( \frac{r}{s}\right)\;.
\end{equation}
Hence,
\begin{equation}
{\tilde I_1}(s) \simeq \frac{\Gamma_d}{ds} \frac{1}{\alpha_0^d}\left[
 \ln \left( \frac{r}{s}\right)\right]^d\;.
\end{equation}
Inverting the Laplace transform one then obtains the large time asymptotic behaviour of
$I_1(t)$
\begin{equation}
I_1(t) \simeq
\frac{\Gamma_d}{d\alpha_0^D}\left[ \ln t \right]^d\;.
\end{equation}
and leading large time behaviour
\begin{equation}
P_s^{\rm av}(t) \sim
\exp\left[ -\rho \frac{\Gamma_d}{d\alpha_0^d}\left( \ln t \right)^d\right]\;.
\label{psdasym}
\end{equation}
Expression (\ref{psdasym}) is the main result of this section.
As a check we can take $d=1$ to recover the result \cite{EM1}
\begin{equation}
P_s^{\rm av}(t) \sim
\exp\left[ -2 \rho \alpha_0 \ln t\right]
=  t^{-2\rho (D/r)^{1/2}}\;.
\label{psd1asym}
\end{equation}
Expression (\ref{psdasym})  shows how the power law decay 
displayed in $d=1$ (\ref{psdasym}) is generalized in arbitrary spatial
dimension. The scaling form $\exp\left[ -A(\ln t)^d \right]$ is unusual,  as far as we are aware.

\subsection{Typical behaviour}
For  the typical behaviour (the quenched case) 
in the long time limit $Q(R,t)$ is dominated by a pole at $s_0$ in the complex $s$ plane (\ref{Qas}), therefore $I_2$, given by (\ref{I2def}), becomes
\begin{equation}
I_2 \simeq 
\mbox{Constant}
- t \Gamma_d \int_a^\infty \D R \, R^{d-1} |s_0(R)|\;.
\label{I2int}
\end{equation}
In general dimension  $d\neq 1$  we were not able to perform the integral
in (\ref{I2int}) explicitly. However it is easy to show that the integral is convergent.
Thus $P_s^{\rm typ}(t)$ given by (\ref{PsI2}) asymptotically decays exponentially in time
\begin{equation}
P_s^{\rm typ}(t)\sim \exp\left[
- t \Gamma_d \int_a^\infty \D R \, R^{d-1} |s_0(R)|\right]\;.
\label{ptypas}
\end{equation}
The correction to the argument of the exponential 
will come from the branch point at $s= -r$ in (\ref{qhdfin})
and is expected to give a subleading contribution of $O(t^{1/2})$. 
The fact that the average and typical survival probabilities 
have distinct asymptotic behaviours---the average behaviour (\ref{psdasym}) decaying far more slowly than the typical behaviour (\ref{ptypas})---reflects the strong  dependence on the initial conditions, noted above, whose memory is retained through resetting.

In dimension  $d=1$ 
the integral in (\ref{I2int}) can be evaluated in closed form.
The reason is that in this case one can obtain a closed form expression for
$R$ as a function of $u_0$ from (\ref{u0eqn})
\begin{equation}
\alpha R = \alpha_0a -\frac{\ln(1-u_0)}{u_0^{1/2}}\;.
\end{equation}
Then one may transform the integration from $R$ to   $u$ with range $ 0 < u <1$ through
$s_0 = r(u -1)$ 
and letting  $\epsilon \to 0$ one finds
\begin{eqnarray}
\int_0^\infty \D R\, |s_0(R)|&=&
\frac{r}{\alpha_0} \int_0^1 \D u \left[
\frac{(1-u)}{u^{3/2}}\ln(1-u)+ \frac{1}{u^{1/2}}\right] \\
&=&
(D r)^{1/2} 4(1-\ln 2)\;.
\end{eqnarray}
Thus in one dimension the  asymptotic decay of the quenched total survival probability is exponential with explicit form  \cite{EM1}
\begin{equation}
P^q_s(t) \sim \exp \left[ - t \rho (D r)^{1/2} 8(1- \ln 2) \right].
\label{Psq}
\end{equation}

\subsection{Explanation in terms of extreme value statistics}
In order to understand the  asymptotic form of average survival probability (\ref{psdasym}),
we consider the following simple picture.  At long times the absorption  probability of the target will be 
dominated by the searcher which started nearest to the target.
Denoting  the position of this searcher by $y$,  the average survival probability
for the many searcher problem  should then be recovered by averaging the  single searcher survival probability (\ref{Qas}) over the distribution of the  position $y$,

In order to obtain the  distribution of the distance $y$ of the nearest
searcher from the   target (at the origin), we consider first the probability that a single searcher starts at distance $R>y$ from the origin
\begin{equation}
\mbox{Prob}( R> y) =\frac{\Gamma_d}{V} \int_y^L \D R  \, R^{d-1}
= \frac{\Gamma_d}{d V}\left[ L^d- y^d\right]\;,
\end{equation}
where $V = \frac{\Gamma_d}{d}\left[ L^d- a^d\right]$\;.

Then the probability that all $N$ searchers start at distance $R>y$ from the origin is given by, in the limit of large $N,L$ with $\rho$ fixed,
\begin{equation}
\mbox{Prob}(R>y)^N  \simeq \exp \left\{ N \left[-\left(\frac{y}{L}\right)^d +
\left(\frac{a}{L}\right)^d \right]\right\}
\to \exp \left[ \rho \frac{\Gamma_d}{d} \left(-y^d + a^d\right)\right]\;.
\end{equation}
Thus the distribution of the distance $y$ of the nearest
searcher from the   target is
\begin{eqnarray}
P(y) &=& - \frac{\D }{\D y }
\exp\left[ \rho \frac{\Gamma_d}{d} \left(-y^d + a^d\right)\right]\\
&=& A\, y^{d-1} \exp \left(
-\rho \frac{\Gamma_d}{d} y^d
\right)\;,
\label{Py}
\end{eqnarray}
where $A$ is a constant.

For a single searcher starting at $y$ the survival probability is given by (\ref{Qas})
\begin{equation}
P_s =  \exp\left[ - t(1-u(y))\right]
\label{Psu}
\end{equation}
where the function $u(y)$ is given by (\ref{u0eqn}).

Thus the average survival probability is given  approximately
by the average of (\ref{Psu}) with respect to (\ref{Py})
\begin{equation}
\langle P_s \rangle
 = A \int_0^\infty \D y\, y^{d-1} \exp\left[-\frac{\rho\Gamma_d}{d}  y^d  - t(1-u(y)) \right]\;.
\label{PSint}
\end{equation}
For large $t$, we expect
the integral to be dominated by the  saddle point $y^*$ of the integral with respect to $y$ which yields
\begin{equation}
-\rho\Gamma_d  (y^*)^{d-1}   +t u'(y^*)=0\;.
\end{equation}
For large $t$ we expect $y^*$ to be  large for which  the asymptotic behaviour is given by (\ref{u0asy})
$u(y) \simeq  1 - \epsilon^{1/2-\nu} 
{\rm e}^{-\alpha_0 y^* (1-\epsilon)}$.
The saddle point $y^*$ is then given by
$ - \rho\Gamma_d (y^*)^{d-1}   + \alpha_0(1-\epsilon)t {\rm e}^{-\alpha_0 y^* (1-\epsilon)} =0$  
which implies that asymptotically
\begin{equation}
y^* 
\sim \frac{\ln t}{\alpha_0}\;.
\end{equation}
Thus at long times $t$ the survival probability is dominated by initial arrangements of searchers  in which  the nearest searcher is at distance $y^* \sim \ln t/\alpha_0$.
The integral (\ref{PSint}) is then dominated by 
\begin{equation}
\langle P_s \rangle
\sim \exp\left[-\frac{\rho\Gamma_d}{d}  (y^*)^d \right]
\sim 
  \exp\left[-\frac{\rho \Gamma_d}{d \alpha_0^d} \left(\ln t\right)^d\right]
\end{equation}
which recovers the asymptotic result (\ref{psdasym}) 
of Subsection \ref{sec:Psa}.

\section{Conclusion}

In this work we have studied diffusion with resetting at rate $r$ in
arbitrary spatial dimension $d$. We have computed the nonequilibrium
stationary state of the process given by Equation (\ref{pssgen}),
which exhibits non-Gaussian behaviour.  Moreover the full
time-dependence is given by (\ref{pt}).  The resetting paradigm thus
presents a simple, general framework in which to study nonequilibrium
properties. We have shown in this work that the spatial dimensionality produces
some interesting behaviour.

In Section 3 we considered the survival probability of an absorbing
target.  Our central result is to compute the Laplace transform of the
survival probability from which long time asymptotic behaviour may be
deduced.  Then in Section 4 we considered the mean time to absorption
of the target. The mean time to absorption is finite and has a minimum
value at an optimal resetting rate $r^∗$.

Finally in Section 5 we consider the problem of a finite density of diffusive particles, each resetting to its own initial position.
While the typical survival probability of the target at the origin  decays exponentially with time regardless of spatial dimension, the average survival probability decays asymptotically as  $ \exp(-A (\ln t)^d)$ where $A$ is a constant. We  have explain these findings using an interpretation as a renewal process and arguments invoking extreme value statistics.

There are many open questions concerning the resetting paradigm. In this work we have considered a single particle system under diffusion but one could
generalise to other stochastic processes, for example, the Ornstein-Uhlenbeck process describing diffusion in a potential. Moreover one can  
generalise resetting to  extended systems where the configuration of the system is reset to the initial condition at a constant rate.
These dynamics generate a new class of nonequilibrium stationary states. Recent examples
that have been considered include
a class of reaction-diffusion systems in one dimension \cite{DHP14} and also  growth processes described by macroscopic  stochastic differential equations such as  the Kardar-Parisi-Zhang and Edwards-Wilkinson equations  \cite{GMS13}.
Naturally, studying the resetting dynamics in these extended systems in higher dimensions   would be an interesting challenge.\\[2ex]

\noindent{\bf Acknowledgements} MRE would like to acknowledge funding from the EPSRC under grant EP/J007404/1. SNM acknowledges support by ANR grant 2011-BS04-013-01 WALKMAT.

\end{document}